\documentclass[copyright,creativecommons]{eptcs}
\providecommand{\event}{PLACE 2022} 

 \usepackage{underscore}           

\usepackage{stmaryrd}
\usepackage{amsmath}
\usepackage{amssymb}
\usepackage{listings}
\usepackage{mathpartir}

\usepackage[sort]{cite} 
\usepackage[T1]{fontenc}

\usepackage{tikz}
\usepackage{lscape}
\usepackage{tikz-cd}
\usetikzlibrary{shapes,arrows,calc,trees,positioning}
\usetikzlibrary{chains,fit,shapes.geometric}
\usetikzlibrary{arrows}

\newcommand{\lnc}{\textsc{Lang-n-Change}}

\newcommand{\quoting}[1]{``#1''}

\newcommand{\app}{\;}
\newcommand{\key}[1]{\ensuremath{\mathtt{#1}}}

\newcommand{\head}{\key{head}}
\newcommand{\tail}{\key{tail}}

\newcommand{\labeledStep}[1]{\step^{#1}}

\definecolor{magenta(dye)}{rgb}{0.79, 0.08, 0.48}
\definecolor{bondiblue}{rgb}{0.0, 0.58, 0.71}
\definecolor{navyblue}{rgb}{0.0, 0.0, 0.5}
\definecolor{lightskyblue}{rgb}{0.53, 0.81, 0.98}

\newcommand*\colourcheck[1]{%
  \expandafter\newcommand\csname #1check\endcsname{\textcolor{#1}{\text{\ding{52}}}}%
}
\colourcheck{blue}
\colourcheck{green}
\colourcheck{red}

\usepackage{mathtools}
\usepackage{relsize}

\newcommand{\ninference}[3]{\inferrule[(#1)]{#2}{#3}}

\newcommand{\ba}{\begin{array}}
\newcommand{\ea}{\end{array}}

\newenvironment{syntax}{\[\ba{l@{\;\;}lcl}}{\ea\]}

\newcommand{\typeOf}{\vdash}
\newcommand{\step}{\longrightarrow}

\definecolor{lightblue}{rgb}{0.25,0.25,1}

\definecolor{lightgray}{gray}{0.9}
\definecolor{darkergrey}{rgb}{0.75, 0.75, 0.75}

\newcommand{\getRulesLNC}{\key{rules}}
\newcommand{\getPremisesLNC}{\key{premises}}
\newcommand{\getConclusionLNC}{\key{conclusion}}

\newcommand{\getVars}[1]{\key{getVars}(#1)}
\newcommand{\isVar}[1]{\key{isVar}(#1)}

\newcommand{\createMapLNC}[2]{\key{map}(#1,#2)}

\newcommand{\GammaForRule}[1]{\Gamma_{\textsf{rule}}}

\newcommand{\andLNC}[2]{#1 \app \key{and}\app #2}
\newcommand{\orLNC}[2]{#1 \app \key{or}\app #2}
\newcommand{\notLNC}[1]{\key{not}\app #1}

\newcommand{\uniquefy}[6]{\key{uniquefy}(#1) \Rightarrow (#4,#5):#6 }

\newcommand{\skipLNC}{\key{skip}}

\newcommand{\msg}[1]{\texttt{\textit{#1}}}

\newcommand{\select}[3]{{#1}[#2]:\app #3}

\newcommand{\errorLNC}{\key{error}}

\newcommand{\ifLang}[3]{\key{if} \app #1 \app \key{then} \app #2 \app \key{else}\app #3}
\newcommand{\ifLangNoELSE}[3]{\key{if} \app #1 \app \key{then} \app #2}

\usepackage{semantic}
\usepackage{centernot}

\definecolor{dkgreen}{rgb}{0,0.6,0}
\definecolor{gray}{rgb}{0.5,0.5,0.5}
\definecolor{mauve}{rgb}{0.58,0,0.82}
\title{A Declarative Validator for GSOS Languages}
\author{Matteo Cimini
\institute{University of Massachusetts Lowell \\ Lowell, MA, USA}
\email{matteo\_cimini@uml.edu}
}
\def\titlerunning{A Declarative Validator for GSOS Languages}
\def\authorrunning{M. Cimini}
\begin{document}
\maketitle

\begin{abstract}
Rule formats can quickly establish meta-theoretic properties of process algebras. 
It is then desirable to identify domain-specific languages (DSLs) that can easily express rule formats. 
In prior work, we have developed $\lnc$, a DSL that includes convenient features for browsing language definitions and retrieving information from them.
In this paper, we use $\lnc$ to write a validator for the GSOS rule format, and we augment $\lnc$ with suitable macros on our way to do so. 
Our GSOS validator is concise, and amounts to a few lines of code. 
We have used it to validate several concurrency operators as adhering to the GSOS format. 
Moreover, our code expresses the restrictions of the format declaratively. 
\end{abstract}

\section{Introduction}\label{intro}

After creating a process algebra, the job of a language designer is not finished yet. 
Ideally, the language designer would strive 
to prove that the process algebra at hand affords 
the desired properties. 
Depending on the process algebra, it may be interesting to establish whether bisimilarity is a congruence, whether the language is deterministic, or whether some of its operators satisfy certain algebraic laws such as commutativity and associativity, to name a few. 

The field of \emph{language validation} aims at developing methods and tools that take a language definition as input, apply a static analysis on it, and establish whether some property holds for the language. In the context of concurrency theory, language validation has been best expressed with \emph{rule formats} \cite{RuleFormats} over Structural Operational Semantics (SOS) specifications \cite{sos}. 
Rule formats state that if the rules that have been 
used to write the SOS specification of a language conform to some syntactic restrictions then some semantic property is guaranteed to hold. 
This approach has been applied to automatically derive the congruence of strong bisimilarity \cite{gsos,tyft,ntyft,panth}, of weak bisimilarity \cite{cool,BLOOM199525,rooted}, and to establish algebraic laws of operators \cite{assoc,zero,commut,distr}, as well as deriving global properties such as determinism \cite{determinism} and bounded nondeterminism \cite{nondet2}, and has also been applied to probabilistic transitions \cite{prob,prob2,prob3} and contexts with binders \cite{bind1,bind2,bind3}, to name a few applications. 

It is then desirable to identify suitable domain-specific languages (DSLs) that make it easier for designers of rule formats to express their formats and automatically test them. 
This allows them to quickly test their new ideas, do so on a suite of several process algebras at once, and have a path to rapidly prototyping new formats. 
These tests are helpful for debugging a new rule format while designers are still crafting a theoretical result. 

Unfortunately, literature does not offer any DSL that has been specifically designed for expressing rule formats. 
In this paper, we focus on an existing DSL called $\lnc$ \cite{lnc1,lnc2}, which has been created for purposes other than language validation but whose operations can be repurposed to write rule formats. 
$\lnc$ is a DSL for expressing language transformations, that is, the input is a language definition and transformation instructions, and the output is a modification of the language given as input. Consider the typing rule of function application below on the left and its version with subtyping on the right. 

$$
\ninference{t-app}
	{
	\Gamma \typeOf \app e_1 : T_1\to T_2 \\
	\Gamma \typeOf \app e_2 : T_1 	
	} 
	{ \Gamma \typeOf \app e_1\app e_2 : T_2}
  \quad \Longrightarrow   \quad
\ninference{t-app'}
	{
	\Gamma \typeOf \app e_1 : T_{11}\to T_2 \\
	\Gamma \typeOf \app e_2 : T_{12} \\  T_{12} <: T_{11}
	} 
	{ \Gamma \typeOf \app e_1\app e_2 : T_2}
$$

\textsc{(t-app')} is a transformation of \textsc{(t-app)}. $\lnc$ provides linguistic features to express such a transformation. 
For example, $\lnc$ can express the test that detects that $T_1$ has been used twice in \textsc{(t-app)} and assigns new distinct variables $T_{11}$ and $T_{12}$ to those occurrences. 
$\lnc$ also includes the operation for creating the premise $T_{12} <: T_{11}$. 
$\lnc$ has been applied to several case studies that automatically transform SOS specifications, such as adding subtyping \cite{lnc1,lnc4}, pattern-matching \cite{lnc2}, references \cite{lnc2}, gradual typing \cite{lnc3}, as well as automatically deriving big-step semantics from small-step style \cite{lnc1}, and CK machines \cite{lnc4}.

To express language transformations, $\lnc$ offers fine-grained operations for browsing SOS rules, their premises, and other components of the language. 
In this paper, we explore the idea of using $\lnc$ to express rule formats. 
The idea is to devise a language transformation that ultimately \emph{does not} modify the language in input, but that uses the operations of $\lnc$ to test the syntactic constraints prescribed by the rule formats, and throw a runtime error when they are not met.  
We have written a checker for the GSOS format \cite{gsos} using $\lnc$. 
This is a well-known rule format, which can establish the congruence of bisimilarity of many process algebras with common operators. 
To better express some of the tests that are prescribed by this format, we have augmented $\lnc$ with suitable macros. 
These are not extensions to the core language, nor to its evaluator. 
These are parsed away into ordinary operations of $\lnc$. 

We have used our tool to validate common concurrency operators that are known to adhere to the GSOS format, such as the CCS parallel operator \cite{ccs}, 
the synchronous parallel composition from CSP \cite{hoare1985communicating}, 
and the projection operator of ACP \cite{acp}. 
(Section \ref{evaluation} provides a complete list of our tests.) 
In total, we have validated 18 concurrency operators. 
We have also performed a series of negative tests. 
More specifically, we have given to our tool languages as input that do not adhere to the GSOS format, 
and we confirm that our tool rejects them, indeed. 

Our GSOS validator amounts to 6 lines of code, which makes for a very concise validator. 
Also, our code expresses the GSOS syntactic restrictions declaratively. 
The work in this paper provides some evidence that $\lnc$ can be a useful tool for expressing rule formats. 

The paper is organized as follows. 
Section \ref{lnc} provides an overview of $\lnc$. 
Section \ref{gsos} presents our new macros and our GSOS validator.  
Section \ref{evaluation} discusses our evaluation. 
Section \ref{relatedWork} discusses related work, and 
Section \ref{conclusion} concludes the paper. 


\section{Overview on $\lnc$}\label{lnc}

We repeat the relevant background on $\lnc$ \cite{lnc1,lnc2} in this section. 
Fig. \ref{pipe} shows the tool pipeline of $\lnc$.
The input consists of two elements: A language definition and a language transformation. The output is either a language definition, or an error message.

 \begin{figure}[tbp]
\begin{center}
\begin{tikzpicture}[
    node distance=2cm,
    auto,
    block/.style={rectangle, draw, fill=cyan!10, text width=6.5cm, text centered, rounded corners, minimum height=4em},
    source/.style={draw,thick,rounded corners,fill=blue!80,inner sep=.2cm},
    tool/.style={draw,thick,fill=blue!20,inner sep=.4cm},
    output/.style={draw,thick,rounded corners,fill=green!50,inner sep=.2cm},
    error/.style={draw,thick,rounded corners,fill=red!70,inner sep=.2cm},
    to/.style={->,>=stealth',shorten >=2pt,ultra thick,font=\sffamily\Huge}]
  \matrix{
   \node[source] (trinput) {\textcolor{yellow}{Language Transformation}}; \\
    \node[source, left of=trinput, xshift=-5cm, yshift=0cm] (laninput) {\textcolor{yellow}{Language Definition}}; \\
    \node[tool, below of=trinput, align=center, yshift=0cm, xshift=-3cm] (lnc) {\large{$\lnc$}}; \\
    \node[output, right of=lnc, yshift=0cm, xshift=-5cm] (lanoutput) {Language Definition}; \\
    \node[error, left of=lnc, xshift=5cm, yshift=2cm] (erroroutput) {Error Message}; \\  
  };
  \draw[to] (trinput) -- (lnc);
  \draw[to] (laninput) -- (lnc);
  \draw[to] (lnc) -- (lanoutput);
  \draw[to] (lnc) -- (erroroutput);
\end{tikzpicture}
\end{center}
\caption{Tool pipeline of $\lnc$}
\label{pipe}
\end{figure}

\paragraph{What Language Definitions?}

$\lnc$ works with languages defined in SOS. The input is a textual representation of transition system specifications for SOS (with negative transitions) \cite{Bol1996}. 
The following is an example language that is input to $\lnc$: A process algebra with the prefix operator, the interleaving operator, and the sequence operator. 

\begin{lstlisting}
Label L ::= (a) | (b) | (c) 
Process P ::= (null) | (a P) | (b P) | (c P) | (par P P) | (sequence P P) 

(a P) --(a)--> P.   

(par P1 P2) --(a)--> (par P1' P2) <== P1 --(a)--> P1'.
(par P1 P2) --(a)--> (par P1 P2') <== P2 --(a)--> P2'.

(sequence P1 P2) --(a)--> (sequence P1' P2) <== P1 --(a)--> P1'.
(sequence P1 P2) --(a)--> P2' <== P2 --(a)--> P2' /\ P1 -/-(a)--> 
                                                        /\ P1 -/-(b)-->
                                                        /\ P1 -/-(c)-->.

(*@ \textit{... rest of the rules (same as the rules above but for the other labels)} @*)
\end{lstlisting}

That is, a grammar declares processes and labels, and a series of inference rules define labeled transitions. 
Intuitively, \textrm{<==} means \quoting{provided that}, and the formulae after that symbol are the premises of the rule. 
A formula such as \quoting{\texttt{P2 -/-(a)--->}} means that $P_2$ does not perform an $a$-transition. 
This syntax does not present any novelty compared to other textual representations for SOS, and is indeed inspired by the syntax employed in the Ott tool \cite{ott}. 

Some remarks on our example language: The GSOS format only works with a finite set of labels \cite{gsos}. 
For simplicity, we have chosen the set of actions $\{a,b,c\}$. Moreover, the uniform setting for operators in GSOS is that of a function symbol applied to processes. This is typically accommodated as shown above: A prefix operator for each action.

\paragraph{What Language Transformations?} 
The following is the subset of the syntax of $\lnc$ that is relevant to this paper\footnote{We refer the reader to \cite{lnc1} and \cite{lnc2} for the syntax of $\lnc$.}. 

 \begin{syntax}
   \text{\sf Expression} & e & ::= & 
   x \mid \textit{str} \mid  t \mid [e \app \ldots \app e]\mid \head \app e \mid \tail \app e \mid e @ e \mid e - e  \mid \createMapLNC{e}{e} \mid  e(e)\\
    &&&\mid \getRulesLNC\mid \getPremisesLNC \mid \getConclusionLNC \mid  \key{self}\\
   &&& \mid \select{e}{p}{e}      \mid \uniquefy{e}{e}{str}{x}{x}{e} \mid \getVars{e}\\  
   &&& \mid \ifLang{b}{e}{e} \mid   \skipLNC  \mid \errorLNC \app str\\
   \text{\sf Boolean Expr.} & b & ::= &  e = e \mid \isVar{e}\mid \andLNC{b}{b} \mid \orLNC{b}{b} \mid \notLNC{b}  \\ 
   \text{\sf Pattern} & p & ::= & x \mid [p \app \ldots \app p]  \mid predname \app p \mid opname \app p \mid x \app p 
   \end{syntax}

We assume a set of operator names \textsc{OpName} ranged over by \emph{opname}. 
\textsc{OpName} contains elements such as \key{par}, and \key{sequence}, for example. 
We also assume a set of predicate names \textsc{PredName} ranged over by  \emph{predname}. 
\textsc{PredName} contains elements such as $\step$ and $\centernot\step$. $\lnc$ accommodates formulae uniformly in abstract syntax $(predname \app \mathit{arg}_1 \app \ldots \app \mathit{arg}_n)$, as it does not make assumptions on the language. Yet, the tool still reads transitions such as $P \labeledStep{a} P'$ with syntax {\texttt{P ---(a)---> P'}} for the convenience of users (see the example language above).

\emph{Expression} is the main syntactic category. 
Given an SOS specification, i.e., a language $\mathcal{L}$, and given an expression $e$, $e$ contains the operations that will be applied to $\mathcal{L}$. 
Expressions can be variables, strings (\textit{str}), terms ($t$) (such as \texttt{(par P1 P2)} and \texttt{(sequence P1 P2)}), and lists with ordinary operations for extracting the head and the tail of lists, appending two lists ($e~@~e$), and performing list difference ($e\;-\;e$). 
Expressions can also be maps $\createMapLNC{e_1}{e_2}$, where $e_1$ and $e_2$ are lists. The first element of $e_1$ is the key of the first element of $e_2$, and so on for the rest of the elements\footnote{This schema is motivated in \cite{lnc1}. For example, it quickly maps $T_i$ to $T_i'$ from conclusions of subtyping rules such as the conclusions $T_1 \to T_2 <: T_1' \to T_2'$ and $T_1 \times T_2 <: T_1' \times T_2'$.}. 
Given a map $m$, $m(k)$ retrieves the value in $m$ associated with the key $k$. 

$\lnc$ includes the special keywords \getRulesLNC, \getPremisesLNC, \getConclusionLNC, and \key{self}. 
The keyword $\getRulesLNC$ returns a list with all the inference rules of the language given as input. 
We shall describe the other keywords in the context of the following operator. 

The \emph{selector operator} $\select{e_1}{p}{e_2} $ selects one by one the elements of the list $e_1$ that satisfy the pattern $p$ and executes the body $e_2$ for each of them. 
The selector returns a list with the values produced by each evaluation of $e_2$. 
The keyword \key{self}, when used in $e_2$, returns the element of the list $e_1$ that has been selected at that iteration. 
A \emph{pattern} $p$ can be a variable, can attempt to match a list (pattern $[p \app \ldots \app p]$), to match a formula that uses a specific predicate name (pattern $\textit{predname} \app p$), to match a term with a specific top-level operator (pattern $\textit{opname} \app p$), or can attempt to match a formula or term with an unspecified top-level name (pattern $x\app p$). 
As typical with pattern-matching, the variables that are used in the pattern $p$ are bound in $e_2$, and are instantiated at runtime. 
To make an example, let us consider $\select{\getRulesLNC}{P \labeledStep{L} P'}{e_2}$ being executed for the example language above (with prefix, \key{par}, and \key{sequence}). 
$\getRulesLNC$ evaluates to a list with all the rules of the language. 
When the list contains rules, the pattern of the selector is applied to match the conclusions of these rules. 
Therefore, the pattern $P \labeledStep{L} P'$ selects all the reduction rules. The first iteration of $e_2$ is executed with $P = \texttt{(a P)}$, $L = \texttt{(a)}$, and $P' = \texttt{P}$, and so on. (Notice that there is no clash between pattern variables and the metavariables of rules, as they are separate in $\lnc$.) 
For the convenience of programmers, simply writing $\select{\getRulesLNC}{\step}{e}$ selects the rules that define $\step$ without specifying a full pattern. 
Also, $e[p]$ is a shorthand for $e[p]:\key{self}$, i.e., a list of the elements selected by the pattern. Therefore, $\getRulesLNC[\step]$ simply selects all reduction rules. 
The keyword $\getPremisesLNC$ can be used when the selector operator works on rules, and returns the list of premises of the selected rule. 
For example, $\select{\getRulesLNC}{\step}{\getPremisesLNC}$ returns a list where each element is the list of premises of a rule such as $[[], [\texttt{P1 ---(a)---> P1'}], [\texttt{P2 ---(a)---> P2'}], \ldots]$ in our example process algebra. 
Similarly, the keyword $\getConclusionLNC$ returns the conclusion of the selected rule.

$\uniquefy{e_1}{e_2}{str}{x}{y}{e_2}$ takes a list $e_1$ of formulae, and returns a version of these formulae where multiple occurrences of a metavariable have been assigned distinct metavariables. 
The computation continues by executing $e_2$. 
The list of new formulae is passed to $e_2$ as $x$. 
\key{uniquefy} also computes a map that summarizes the changes that have been made to the original list of formulae $e_1$. 
This map is passed to $e_2$ as $y$. This operation is useful for transformations such as that of \textsc{(t-app)} into \textsc{(t-app')} that we have described in Section \ref{intro}. 
Suppose that $l$ contains the list of premises of \textsc{(t-app)}, then $\uniquefy{l}{e_2}{str}{\emph{newPremises}}{\emph{mapOfChanges}}{e}$ will execute $e$ where \emph{newPremises} is the list $[\Gamma \typeOf \app e_1 : T_{11}\to T_2,  \Gamma \typeOf \app e_2 : T_{12}]$ and $\emph{mapOfChanges}$ is the map $\{T_1\mapsto [T_{11}, T_{12}]\}$, which denotes that the occurrences of $T_1$ have been split into $T_{11}$ and $T_{12}$\footnote{\key{uniquefy} can be used in a more fine-grained style, as shown in \cite{lnc1}, but we do not need that style in this paper.}.

$\getVars{e}$ returns the list of metavariables that are used in $e$ after it has been evaluated. 
We also have an if-statement, and a $\skipLNC$ operation that does not perform any operation. 
When \key{if} has no \key{else} branch, as in $\ifLangNoELSE{b}{e}{e}$, it means $\ifLang{b}{e}{\skipLNC}$. 
Error $\errorLNC$ throws a runtime error and carries a string as error message. 
The boolean conditions of the if-statement can check for syntactic equality, whether $e$ is a metavariable with $\isVar{e}$, and can combine these checks with boolean operations. 

We do not discuss here the type checker of $\lnc$, which has been presented in \cite{lnc1} and can reject, for example, $\select{e}{p}{e'}$ when $e$ is not a list, and other type errors.  

\section{A GSOS Validator}\label{gsos}

\paragraph{The GSOS Rule Format}

We recall the GSOS format \cite{gsos}. 
The following is the shape for GSOS rules: 

\[
\inference
{\{x_i \step^{l_{ij}} y_{ij} \mid i \in I, 1\leq j \leq m_i\}
~ \cup ~ 
\{x_j \centernot\step^{l'_{jk}}  \mid j \in J, 1\leq k \leq n_j\}
}
{(op \app x_1 \ldots x_h) \step^{l} t}
\]

Notice that $x$s and $y$s are metavariables for metavariables, so that some relation can be stated among different metavariables. 
In other words, $x$s and $y$s all denote metavariables such as $P$, $P_1$, $P_2$, and so on. 
We have that $x_i$ and $y_i$ are all distinct. 
$I$ and $J$ are subsets of $\{1, \dots, h\}$, that is, $x$s in the premises come from the conclusion, and each of them can be the subject of positive premises multiple times, as well as the subject of negative premises multiple times. 
The metavariables that occur in $t$ can only come from $x$s and $y$s, 
Finally, labels $l$s are constants. 

A rule that conforms to these restrictions is a GSOS rule. 
For example, all the rules of the example process algebra in the previous section (with prefix, \key{par}, and \key{sequence}) are GSOS rules. 

Consider the following rule, which defines the behavior of the replication operator. 

$$
\qquad\quad
\inference
{(P \mid \; !P) \labeledStep{a} P'}{!P \labeledStep{a} P'}
$$

This rule is not a GSOS rule because the source of the premise is $(P \mid \; !P)$ rather than a variable. 

The following is a classic result of the meta-theory of SOS: If all the rules of the language are GSOS rules then bisimilarity is a congruence for the language \cite{gsos}. 

\subsection{New Macros for $\lnc$} 
We define the following macros in $\lnc$. 

\begin{itemize}

\item $e \app \key{must\app match} \app p_1 \mid p_2 \mid \ldots \mid p_n \app \key{otherwise}\app  e' \triangleq \key{if} \app \key{not}((e - e[p_1] - e[p_2] \ldots - e[p_n]) = [])\app \key{then} \app e'$. 

Here, $e$ is a list. The idea is that each element of $e$ must match one of the patterns $p_1, p_2, \ldots, p_n$, otherwise we execute $e'$. 
To do that, we progressively subtract from $e$ its sublists filtered by the patterns and check that the resulting list is empty. 
This macro is useful to check that premises and conclusions have the correct shape.  
We use this same \quoting{empty list}-test to check whether a list is a sublist of another with: $e \app \key{sublistOf} \app e' \triangleq (e - e') = []$. 
This macro is useful to check that the metavariables being used in some part of the rule all come from the correct list of metavariables. 

\item $\key{match} \app e\app \key{with} \app p \to e'  \app \key{otherwise}\app e'' \triangleq \key{if} \app [e][p] = [] \app \key{then} \app e'' \app \key{else}\app e'$. 

Here, we check that $e$ matches the pattern $p$ and, if that is the case, we execute $e'$, otherwise we execute $e''$. To do so, we create the list with only one element $[e]$ and use the selector to filter it by pattern $p$. If the resulting list is empty then the pattern $p$ does not succeed for $e$\footnote{\cite{lnc3} used this method for a simpler version of this macro.}. 
When we omit \quoting{$\to e'$} in this macro, it means \quoting{$\to \key{skip}$}. 
When we omit \quoting{$\key{otherwise}\app e''$}, it means \quoting{$\key{otherwise}\app \key{skip}$}. 
This macro is useful to check a pattern for one element, as opposed to a list as above, and to specify a \key{then}- versus \key{otherwise}-reaction.

\item The following macros are useful to quickly access sources and targets of transition formulae: 
\begin{align*}
\key{premises.LTsources} & ~\triangleq~ (\key{premises[\text{\texttt{$P\;$---$L$--->$\;P'$]:$P$}}}) ~@~ (\key{premises[\text{\texttt{$P\;$--/-$L$--->]:$P$}}})\\[-.7ex]
\key{premises.LTtargets} &~\triangleq  ~\key{premises[\text{\texttt{$P\;$---$L$--->$\;P'$]:$P'$}}}\\[-.7ex]
\key{conclusion.LTsource} &~\triangleq ~\key{head ~ ([conclusion][\text{\texttt{$P\;$---$L$--->$\;P'$]:$P$}}})\\[-.7ex]
\key{conclusion.LTtarget} & ~\triangleq  ~\key{head ~ ([conclusion][\text{\texttt{$P\;$---$L$--->$\;P'$]:$P'$}}})
\end{align*}

Notice that \key{premises.LTsources} extracts the sources of both positive and negative labeled transition formulae. 
\quoting{\key{LT}} in \key{LTsources} stands for labeled transition. 
We also have introduced the analogous macros for (unlabeled) transitions $P\step P$ such as \key{Tsources}, \key{Ttargets}, and so on. 
We do not think that these are ad-hoc macros in the context of language design. Labeled and (unlabeled) transitions are so common that it is reasonable to have operations that say, for example, \quoting{\emph{handle this premise as a labeled transition formula and return its source}}. 
When a formula of another shape is given, \key{premises.LTsources} and \key{premises.LTtargets} return an empty list, and \key{head} fails at runtime for \key{conclusion.LTsource} and \key{conclusion.LTtarget}. 

\item $\key{distinctVars}(e) \app \key{otherwise}\app e' \triangleq$

\qquad$\uniquefy{[(\emph{pname} \app e)]}{e_2}{str}{\emph{new}}{\emph{m}}{\key{if} \app \key{not}(\emph{m} = \createMapLNC{[]}{[]}) \app \key{then}\app e'} $ 

Here, $e$ is a list of metavariables. We create the formula $(\emph{pname} \app e)$ with an unused predicate name \emph{pname} just so we can pass it to \key{uniquefy}. 
If $m$ is the empty map $\createMapLNC{[]}{[]}$ then \key{uniquefy} did not detect any metavariable as being used more than once, i.e., all metavariables in $e$ are distinct. 
\key{distinctVars} executes $e'$ otherwise. 

\end{itemize}

\subsection{A GSOS Validator in $\lnc$}\label{checks}

We now use $\lnc$ to write a GSOS validator. We divide our task into $5$ parts. 
These are $5$ checks that are meant to be performed in the order they appear, i.e., first Part 1, then Part 2, and so on. 
They all return $\skipLNC$ if their corresponding check succeeds, otherwise they throw a runtime error. 
Thanks to the operations of $\lnc$ and our macros, these checks are easy to read, and we may omit commenting on some of them. 
Below, we use $\mathit{math}$ font for $\lnc$ pattern variables. 

\begin{itemize}
\item Part 1: All premises are positive or negative transition formulae, and they use constant labels. 

\begin{lstlisting}
rules[-->]: premises must match (*@$\mathit{P}\;$@*)--((*@$\mathit{op}$@*) [])-->(*@$\;\mathit{P'}$@*) | (*@$\mathit{P}\;$@*)-/-((*@$\mathit{op}$@*) [])--> 
             otherwise error (*@\msg{msg}@*)
\end{lstlisting}

where $\msg{msg} =$ \quoting{Premises must be either positive labeled transitions or negative labeled transitions, and their label must be a constant}. 
A constant is a term with a top-level operator and an empty list as arguments. 

\item Part 2: All conclusions are transition formulae that use a constant label, and are defined for an operator applied to metavariables as arguments. (Part 4 will check later that these metavariables are distinct, as they also need to be distinct from $y$s.)  

\begin{lstlisting}
rules[-->]: match conclusion with ((*@$\mathit{op_1}~Ps\;$@*))--((*@$\mathit{op_2}$@*) [])-->(*@$\;\mathit{P'}$@*) -> 
                   (*@$\mathit{Ps}$@*)[(*@$\mathit{P}$@*)]: if not(isVar((*@$\mathit{P}$@*))) then error (*@\msg{msg}$_1$@*) 
             otherwise error (*@\msg{msg}$_2$@*) 
\end{lstlisting}

where $\msg{msg}_1 =$ \quoting{The operator that is the subject of the conclusion must have all metavariables as arguments}, and $\msg{msg}_2 =$ \quoting{Conclusion formulae must be positive labeled transitions with a constant label and must apply to an operator}. 

\item Part 3: Sources of premises must come from $x$s of the conclusion, and $y$s must be metavariables.  

\begin{lstlisting}
rules[-->]: 
  if not(premises.LTsources (*@\isContainedIn@*) getVars(conclusion.LTsource)) 
  then error (*@\msg{msg}$_1$@*) 
  else premises.LTtargets[(*@$\mathit{P}$@*)]: if not(isVar((*@$\mathit{P}$@*))) then error (*@\msg{msg}$_2$@*) 
\end{lstlisting}

where $\msg{msg}_1 =$ \quoting{Sources of premises must be arguments of the operator in the source of the conclusion}, 
and $\msg{msg}_2 =$ \quoting{Targets of premises must be metavariables}. 
Here, \key{premises.LTsources}, \key{conclusion.LTsource}, and \key{premises.LTtargets} are used after Part 1 and Part 2 have checked that we do have labeled transition formulae. 

\item Part 4: $x$s in the source of the conclusion, and $ys$ in the premises must all be distinct. 

\begin{lstlisting}
rules[-->]: 
  distinctVars (getVars(conclusion.LTsource) @ premises.LTtargets) 
  otherwise error (*@\msg{msg}@*)
\end{lstlisting}

where $\msg{msg} =$ \quoting{The arguments of the operator in the source of the conclusion and the targets of the premises must all be distinct metavariables}.

\item Part 5: Metavariables in the target of the conclusion come from $x$s and $y$s.

\begin{lstlisting}
rules[-->]:     
 if not(getVars(conclusion.LTtarget) 
                (*@\isContainedIn@*) 
         (getVars(conclusion.LTsource) @ premises.LTtargets)) 
 then error (*@\msg{msg}@*)
 \end{lstlisting}

where $\msg{msg} =$ \quoting{The metavariables in the target of the conclusion must come from the source of the conclusion or from the targets of premises}. 

\end{itemize}

\section{Evaluation}\label{evaluation}

We have implemented the macros that we have described in this paper \cite{tool}\footnote{The flagship implementation of $\lnc$ is that of \cite{lnc2} but it uses a syntax that is more verbose than the one firstly proposed in \cite{lnc1}. We have then implemented a lightweight evaluator of the $\lnc$ DSL. The flagship implementation of $\lnc$ remains that of \cite{lnc2}.}. 
More precisely, we have not changed the core language of $\lnc$, 
but we have added those constructors to the surface language for the convenience of programmers. 
These constructors are simply \emph{parsed away}. 

We have created a collection of test cases for our GSOS validator. 
We have defined a base language definition with only the prefix operator $l.P$. This language serves as a base to which we have added other features. 
Starting from this, we have created one language for each of the following concurrency operators: 
the interleaving parallel operator of CCS \cite{ccs} (without process communication), 
the full parallel operator with communication of CCS \cite{ccs}, 
the synchronous parallel composition from CSP \cite{hoare1985communicating}, 
the external choice of CCS,
the internal choice of CSP, 
projection of ACP \cite{acp}, 
hiding of CSP, 
left merge operator, 
the rename operator of CCS, 
the restriction operator of CCS, 
the \quoting{hourglass} operator  from \cite{hourglass}, 
signaling \cite{signaling}, 
the disrupt operator, 
the interrupt operator, 
the sequence operator $;$, 
the priority operator, 
and a while-loop operator.  
These operators are known to satisfy the GSOS restrictions. 

Our repo contains 18 concurrency operators, and we confirm that our GSOS validator successfully executes the checks of Section \ref{checks} (Part 1--5) on all these languages. 
That is, our tool validates the above-mentioned operators as adhering to the GSOS format.  
We have also performed a series of negative tests. 
Specifically, we have created languages that do not conform to the GSOS restrictions described in Section \ref{gsos}. 
We confirm that our GSOS validator fails in these cases and provide the corresponding error message.  

Most of the checks of Section \ref{checks} can be written in one line despite having presented them in multiple lines for readability. 
Overall, we could write a GSOS validator in 6 lines of $\lnc$ code. This is a remarkably concise validator. 
Moreover, we believe that our code expresses the syntactic restrictions of the GSOS format declaratively. 
The website of our tool reports on all our tests \cite{tool}.

\section{Related Work}\label{relatedWork}

We are not aware of domain-specific languages that have been designed to express rule formats. 


There are only a few tools that validate rule formats. 
\textsf{PREG Axiomatizer} \cite{Goriac2} includes a checker for the GSOS format in around four hundred lines of Maude code\footnote{We are thankful to Eugen{-}Ioan Goriac for kindly providing this estimation in private communications via email, and also for clarifying that such a GSOS checker is originally a part of \textsf{PREG Axiomatizer} rather than \textsf{Meta SOS}. Notice that if we did not count the code for parsing, the checker is estimated to be around 150 lines of Maude code.}. 
Besides the format checks, this part implements methods for retrieving information from languages. For example, it implements functions for retrieving rules, searching premises, and obtaining the variables used in formulae, to name a few. As it turns out, these are functionalities that most language tools \cite{verifier,gradualizer,Grewe:2015,lnck,Roberson} need to use. 
It appears that each tool makes use of a specific programming language, stores languages as a data type of such programming language, and reimplements these retrieval functions. 
This is an issue that $\lnc$ can alleviate by providing a DSL for expressing them concisely and declaratively. 
It would be interesting to embed $\lnc$ into programming languages so that implementors can call its primitives. 

\textsf{Meta SOS} \cite{Goriac} implements rule formats other than the GSOS format, hence a direct comparison is not possible. 
The tool of Mousavi and Reniers \cite{MousaviMaude} provides a GSOS validator in Maude, and adopts a different implementation approach than \cite{Goriac2}. 
Process algebras are provided as Maude rewriting rules. 
The tool then makes use of Maude introspective reflective features to explore the shape of rules, premises, and so on. 
Language designers can certainly use this approach to express their next rule formats, but it requires familiarity with Maude, with its reflective library, and with a very particular style of meta-programming that can have a steep learning curve. 
Some practitioners may find the linguistic features of $\lnc$ more intuitive and accessible.


\section{Conclusion}\label{conclusion}
Rule formats can quickly establish meta-theoretic properties of process algebras. 
It is then desirable to identify DSLs that can easily express rule formats and automatically test them. 
In this paper, we have observed that $\lnc$ offers convenient operations to interrogate operational semantics. 
We have created macros on top of $\lnc$ to better express some of the checks that often occur in rule formats. 
We have then used $\lnc$ and our macros to implement the GSOS rule format. 
Overall, we have written a full GSOS validator with only 6 lines of code, and we have used it to validate several concurrency operators. 
Moreover, our code expresses the GSOS restrictions declaratively. 

In the future, we would like to apply our approach to other rule formats \cite{RuleFormats}. 
Several formats, including tyft, ntyft, path and panth \cite{tyft,ntyft,panth}, check for distinct variables, impose a specific shape for transition formulae, and retrieve sources and targets for analysis. 
We believe that $\lnc$ and our macros can be useful in those cases. 
However, there are possibly challenging aspects of our approach. 
For example, there may be operations that $\lnc$ does not implement yet, and whose need may be discovered at the attempt of capturing other rule formats. 
We have encountered an instance of this scenario when using $\lnc$ to automatically add references to certain pure functional languages \cite{lnc2}. 
This endeavor requires lifting the shape of reduction rules from $e\step e$ to $e ; \mu \step e ; \mu$, where $\mu$ is the heap. 
In that occasion we have extended $\lnc$ with an operator called \key{lift} to specify the change of shape for relations. 
That is, we have added a new operator that $\lnc$ lacked. (This operator turned out to be useful also for automating gradual typing in \cite{lnc3}, and it is therein described.) 
Another challenge is that $\lnc$ presents some limitations. 
For example, a characteristic of the GSOS rule format is that its checks remain \quoting{local} and confined to the rule that has been selected for analysis.  
Some rule formats, instead, need to maintain a more global view on the process algebra at hand. An example is the rule format that establishes the commutativity of operators (modulo bisimilarity) \cite{commut}, which compares multiple rules at the same time. 
$\lnc$ seems to be best suited to select one rule at a time. 
We will explore developing other DSLs, if other linguistic designs are more suitable.

Our tool is publicly available, and all our tests are documented at its GitHub repo \cite{tool}.

\bibliographystyle{eptcs}
\bibliography{allgsos}

\end{document}